# Indirect Detection of Lactate Through Voltammetry Using Glassy Carbon Microelectrodes

**Amish Rohtagi[a,b], Elisa Barts [a,b], Neharika Ravichandran [a,b], Sandra Lara Galindo[a,b], Surabhi Nimbalkar[a,b], Mantra Mittal [a,b], Samantha Omer[a,b], and Sam Kassegne[a,b,1]**

[a] NanoFAB.SDSU Research Lab., Department of Mechanical Engineering, College of Engineering, San Diego State University, 5500 Campanile Drive, San Diego, CA 92182-1323, USA
[b] Center for Neurotechnology (CNT)

## Abstract

Glassy carbon (GC) microelectrodes are increasingly being used for voltametric detection of electroactive neurotransmitters such as dopamine and serotonin. However, non-electroactive molecules including lactate, glutamate, and gamma-aminobutyric acid (GABA) cannot be directly detected using conventional voltammetry without surface functionalization. In this study, lactate oxidase was immobilized within a chitosan matrix on lithographically patterned GC microelectrodes to enable indirect detection of lactate via enzymatic generation of hydrogen peroxide ($H_2O_2$), an electroactive byproduct. The resulting $H_2O_2$ was detected using fast-scan cyclic voltammetry (FSCV), enabling indirect *in vitro* detection of lactate at concentrations as low as 10 nM. The functionalized GC microelectrodes were integrated into a four-channel array (30 µm × 60 µm sites) on a 1.6 cm flexible neural probe with potential for *in vivo* applications. Surface morphology and bonding interactions were characterized using scanning electron microscopy (SEM) and Fourier-transform infrared (FTIR) spectroscopy. FTIR analysis confirmed successful chitosan deposition through characteristic *O–H, N–H*, *amide*, and *C–O* stretching bands. Hydrogen peroxide detection was concentration-dependent, while lactate detection exhibited early saturation consistent with enzyme-limited kinetics. These results demonstrate a mechanically robust GC microelectrode platform for nanomolar-level indirect lactate sensing and provide insight into the reaction–diffusion coupling governing enzyme-based electrochemical detection.

## 1. Introduction

Real-time monitoring of neurochemical signaling is essential for understanding brain function and dysfunction. Neurotransmitters and signaling molecules regulate critical physiological processes including voluntary movement, reward processing, cognition, and synaptic plasticity. Dysregulation of neurotransmitter release and uptake has been implicated in numerous neurological and psychiatric disorders, including Parkinson's disease, schizophrenia, spinal cord injury, epilepsy, and addiction [**1–8**]. Lactate, the focus of this study, is of particular interest as both a metabolic substrate and signaling molecule within the central nervous system (CNS). It plays a key role in neuronal energy metabolism, ischemic and hypoxic responses, seizure activity, and neuroprotection [**9–15**]. Furthermore, lactate has demonstrated neuroprotective effects in the early stages of certain diseases, whereas its deficiency may contribute to cognitive disorders [**15**]. Accordingly, tools capable of detecting rapid fluctuations in neurochemical concentrations within the CNS are indispensable for both mechanistic studies and the development of closed-loop therapeutic platforms.

---

[1] Address correspondences to Sam Kassegne • Professor of Mechanical Engineering, NanoFAB.SDSU Research Lab, Department of Mechanical Engineering, College of Engineering, San Diego State University, 5500 Campanile Drive, CA 92182-1323. E-mail: kassegne@sdsu.edu • Tel: (760) 402-7162.

From a detection point of view, neurotransmitters and signaling molecules can be broadly categorized as electroactive or non-electroactive. Electroactive species such as dopamine (DA), serotonin (5-HT), and adenosine undergo direct oxidation–reduction reactions at the microelectrode surfaces and can be detected using techniques such as fast-scan cyclic voltammetry (FSCV) through their characteristic redox peaks [**16–20**]. In contrast, non-electroactive molecules, including glutamate, lactate, acetylcholine, and gamma-aminobutyric acid (GABA), do not undergo direct electrochemical oxidation under conventional voltametric conditions and therefore require indirect detection strategies [**21-22**].

As an alternative, due to the minimal requirement for specialized equipment and data post-processing, amperometric biosensors have traditionally been used to detect non-electroactive neurotransmitters *via* enzyme-mediated generation of electroactive products [**23–24**]. However, amperometric approaches often suffer from limited selectivity, slower temporal resolution, and challenges in multiplexed detection. FSCV, by contrast, offers sub-second temporal resolution that matches synaptic timescales and enables simultaneous monitoring of multiple analytes. Recent work has demonstrated enzyme-based indirect detection of lactate using chitosan hydrogels electrodeposited onto carbon-fiber microelectrodes (CFMs) [**25-26**]. While effective, CFMs present limitations including mechanical fragility, inconsistent surface properties, and limited scalability, all of which constrain reproducibility and integration into multi-site neural probe architectures [**27**].

To overcome these limitations, lithographically patternable carbon-based microelectrodes have emerged as promising alternatives. Among these materials, glassy carbon (GC) combines mechanical robustness, microfabrication compatibility, favorable adsorption properties, antifouling behavior, and stable electrochemical performance. GC microelectrodes have demonstrated reliable detection of electroactive neurotransmitters at nanomolar concentrations and compatibility with multi-site neural interfaces [**28–30**]. Recently, these electrodes have also been used for the indirect detection of glutamate, one of the key non-electroactive neurotransmitters, through a wide range of concentrations [**22-23**]. Extending GC platforms to other non-electroactive neurotransmitters such as lactate requires a similar and efficient enzyme immobilization strategies. Indirect electrochemical detection of lactate can be achieved by immobilizing lactate oxidase, which catalyzes conversion of lactate to pyruvate and hydrogen peroxide (L-lactate + $O_2$ → pyruvate + $H_2O_2$), enabling detection of the electroactive $H_2O_2$ product.

In this study, we integrate lactate oxidase within a chitosan matrix onto lithographically patterned GC microelectrodes to enable indirect *in vitro* detection of lactate using FSCV. Beyond demonstrating nanomolar detectability, we systematically evaluate electrochemical transduction, enzyme-limited kinetics, and reaction–diffusion coupling within the immobilized layer. By decoupling electrode performance from enzymatic transport limitations, this work establishes a scalable GC-based platform for multiplexed neurochemical sensing of both electroactive neurotransmitters and metabolically relevant analytes.

## 2. Materials and Methods

### 2.1 Microfabrication of Microelectrodes

A 1.6 cm long penetrating neural probe with four GC microelectrodes of 30 μm x 60 μm size was microfabricated. As described elsewhere, the microfabrication process for the probe shown in **Figure 1a** involved spin-coating of SU-8 negative photoresist (Microchem, MA) on a silicon wafer (with 0.5 μm thick oxide layer) at 1200 rpm for 55 s and soft-baking at first 65°C for 10 min and then 95°C for 20 min followed by UV exposure at ~400 mJ/cm$^2$ [**30-32**]. This was followed by a post-exposure bake at 65°C for 1 min and 95°C for 5 min, development of SU-8 for 3–5 min and curing at 150°C for 30 min. Pyrolysis was done at 1000°C in an inert $N_2$ environment following protocols described elsewhere [**30**]. Subsequently, 5 μm layer of photo-patternable polyimide (HD 4100) (HD Microsystems, DE, USA) was spin-coated on top of the GC microelectrodes at 2500 rpm for 45 s, soft baked at 90°C for 3 min and at 120°C for 3 min, then cooled down to room temperature, and patterned through UV exposure at ~400 mJ/cm$^2$. Then, the polyimide layer was partially cured at 300°C for 60 min under $N_2$ environment. Following, Pt metal traces with Ti adhesion layer were patterned using a metal lift-off process. For electrical insulation, an additional 6 µm of polyimide HD 4100 was spin-coated (300 rpm), patterned (400 mJ/cm$^2$), and cured (350°C for 90 min) under $N_2$ environment. An additional 30 µm thick layer of polyimide (Durimide 7520, Fuji Film, Japan) was spin-coated (800 rpm, 45 s) and then patterned (400 mJ/cm$^2$) on top of the insulation layer to reinforce the penetrating portion of the device (**Figure 1b**). Once the probes were released from the carrier substrate using a BHF bath, the GC microelectrodes were plasma etched (120W for 45 s) and then functionalized through drop-casting of a thin coat of an enzyme mix. The same process was applied to a second set of neural probes, with a length of 1.7 cm and six GC microelectrodes of 40 µm in diameter.

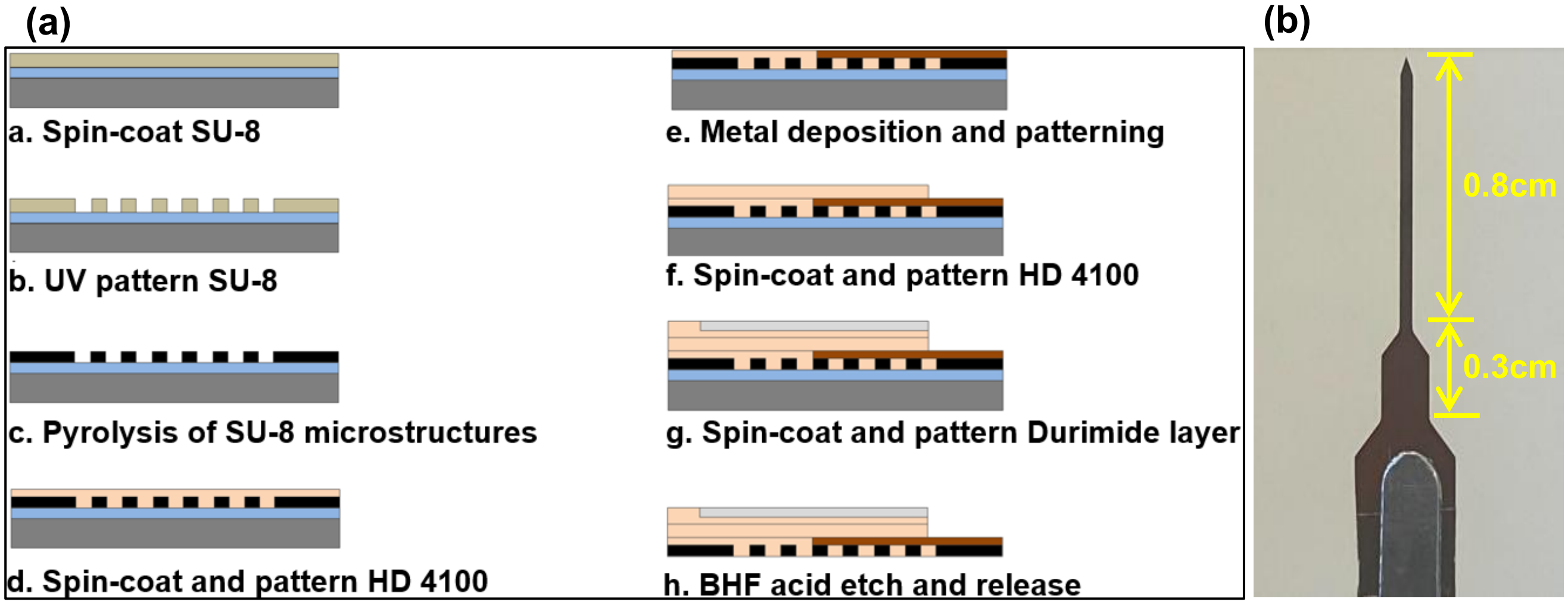

**Figure 1**. (a) microfabrication steps for *4-channel* penetrating neural probes (1.6 cm long) mounted on polymeric substrate (b) fabricated probe for neurotransmitter detection application.

## 2.2 Electrochemical Characterizations

The electrochemical behavior of the microelectrodes was studied in 0.01M PBS (phosphate-buffered saline solution) with pH 7.4 (Sigma Aldrich, USA). Electrochemical impedance spectroscopy (EIS) was used to determine the electrical properties of the probes over a large range of frequencies. using a Potentiostat (Reference 600+, Gamry Instruments, USA) connected to a three-electrode electrochemical cell with a platinum wire as a counter electrode and a saturated Ag/AgCl reference electrode. For EIS measurements, 10 mV RMS amplitude sine wave was superimposed on 0 V potential with frequency sweep from 0.1 to $10^5$ Hz. Equivalent circuit modeling of the EIS data was done through Gamry Echem Analyst Vn 7.05 software (Gamry Instruments, USA).

## 2.3. Functionalization of Microelectrodes

As shown in **Figure 2**, the surface functionalization process consisted of preparing an enzyme immobilization matrix made of *lactate oxidase*, a catalytic enzyme. For that, *Chitosan* (≥ 75% degree of deacetylation) was purchased from Sigma Aldrich (St. Louis, MO) and was diluted in a 0.1 Molar solution of *Sodium Acetate Buffer* to create 20 ml of 0.25% *chitosan* solution A 0.25% (w/v) *chitosan* solution was prepared in 0.1 M sodium acetate buffer. The solution was filtered through Grade 1 Whatman® qualitative filter paper prior to enzyme addition. The cloudy *chitosan acetate* solution was then filtered through Grade 1 Whatman® qualitative filter paper. The enzyme, 1.2 mg of *lactate oxidase* was then dissolved in 100 µl of the newly formed *chitosan* solution. Once the *lactate oxidase* was sufficiently mixed, the microelectrodes were then dipped with 20 µl of the solution with micropipettes.

### 2.4 Scanning Electron Microscopy and FTIR Spectroscopy

Four samples were prepared for scanning electron microscopy (SEM) and Fourier-transform infrared (FTIR) spectroscopy characterization: (1) bare GC, (2) plasma-etched GC, (3) plasma-etched GC with chitosan coating, and (4) GC with chitosan coating. FTIR spectroscopy was performed to identify surface functional groups and, by inference, assess the bonding interactions between the GC microelectrode surface and the immobilized functionalization matrix. Measurements were conducted using a Nicolet iS50 FTIR spectrometer equipped with a Smart iTR diamond attenuated total reflectance (ATR) cell (Thermo Scientific, Vernon Hills, IL, USA). Prior to analysis, each sample was washed with methanol, dried, and placed on the diamond ATR crystal. For each sample, 128 spectral scans were collected and compiled from four diagonal regions of the microelectrode surface.

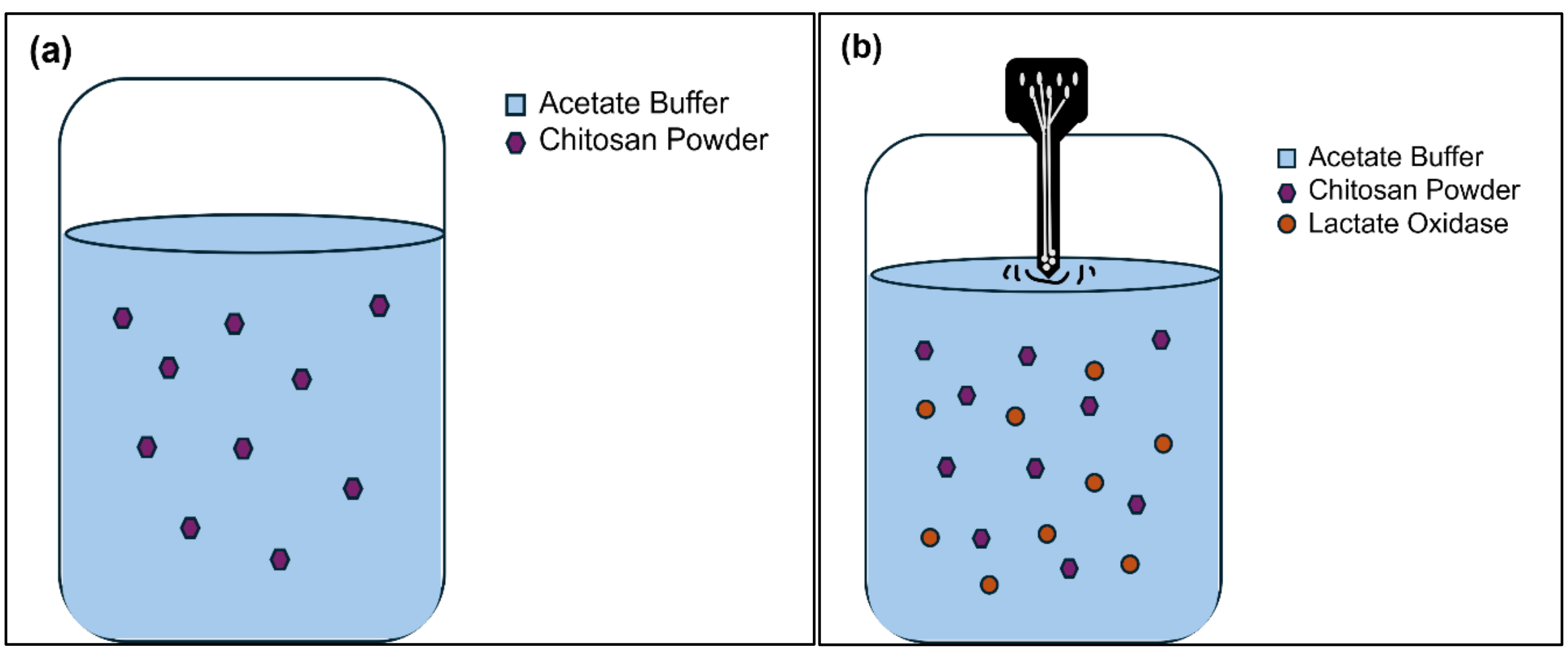

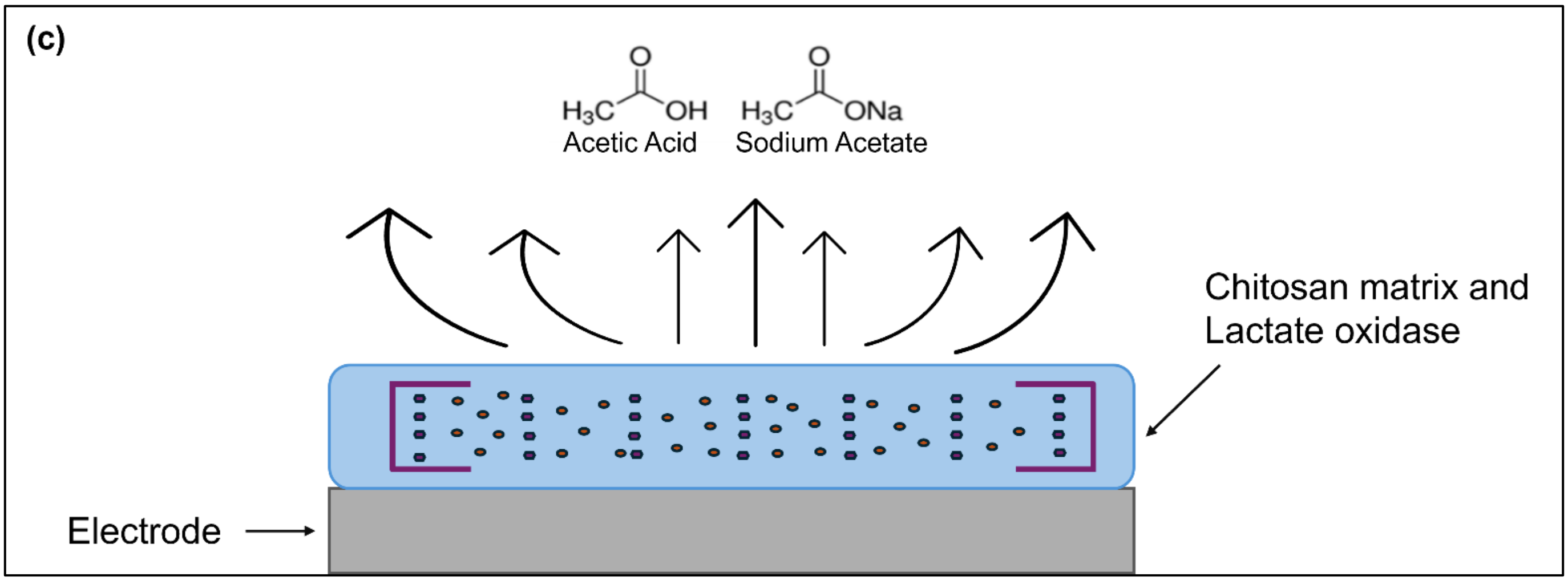

**Figure 2**. (a) *Chitosan* solution (b) *Chitosan* solution with *lactate oxidase* (c) solvent evaporation at microelectrode surface. Example of a neural probe dipped into the enzymatic chitosan solution.

### 2.5 Voltammetry

Indirect detection of lactate *via* direct electrochemical detection of its conversion product, hydrogen peroxide ($H_2O_2$), was performed using a WaveNeuro Potentiostat System (Pine Research, NC). The surface-functionalized GC microelectrode served as the working electrode, with a standard silver/silver chloride (Ag/AgCl) electrode used as the reference. The electrolyte solution consisted of phosphate-buffered saline (PBS) (0.01 M, pH 7.4; Sigma-Aldrich, USA). A triangular waveform was applied at a scan rate of 400 V/s, sweeping from −0.5 V to +1.3 V with a holding potential of −0.5 V versus the Ag/AgCl reference electrode. Each scan had a duration of 9 ms and was repeated at a frequency of 10 Hz. For electrode preconditioning, the same voltage waveform was applied at 60 Hz for 1 hour prior to each experiment.

Known concentrations of lactate (10 nM – 1.6 µM) were infused over 5 seconds, and the resulting current changes were recorded for 30 seconds. As a control, a separate probe without surface functionalization was subjected to the same waveform. Data analysis and background-subtracted cyclic voltammograms (CVs) were processed using HDCV software (UNC Chapel Hill).

## 3. Results

### 3.1 Surface Morphology Characterization

Scanning electron microscopy (SEM) was performed to evaluate morphological changes following chitosan-based enzyme immobilization on the glassy carbon (GC) microelectrodes (**Figure 3**). The bare GC surface appeared smooth and featureless at the micron scale. After drop-casting the chitosan–lactate oxidase matrix, the surface exhibited a porous, mesh-like morphology consistent with polymer film formation. The interconnected structure suggests formation of a continuous coating layer capable of retaining immobilized enzyme while maintaining analyte permeability. No evidence of macroscopic cracking or delamination was observed.

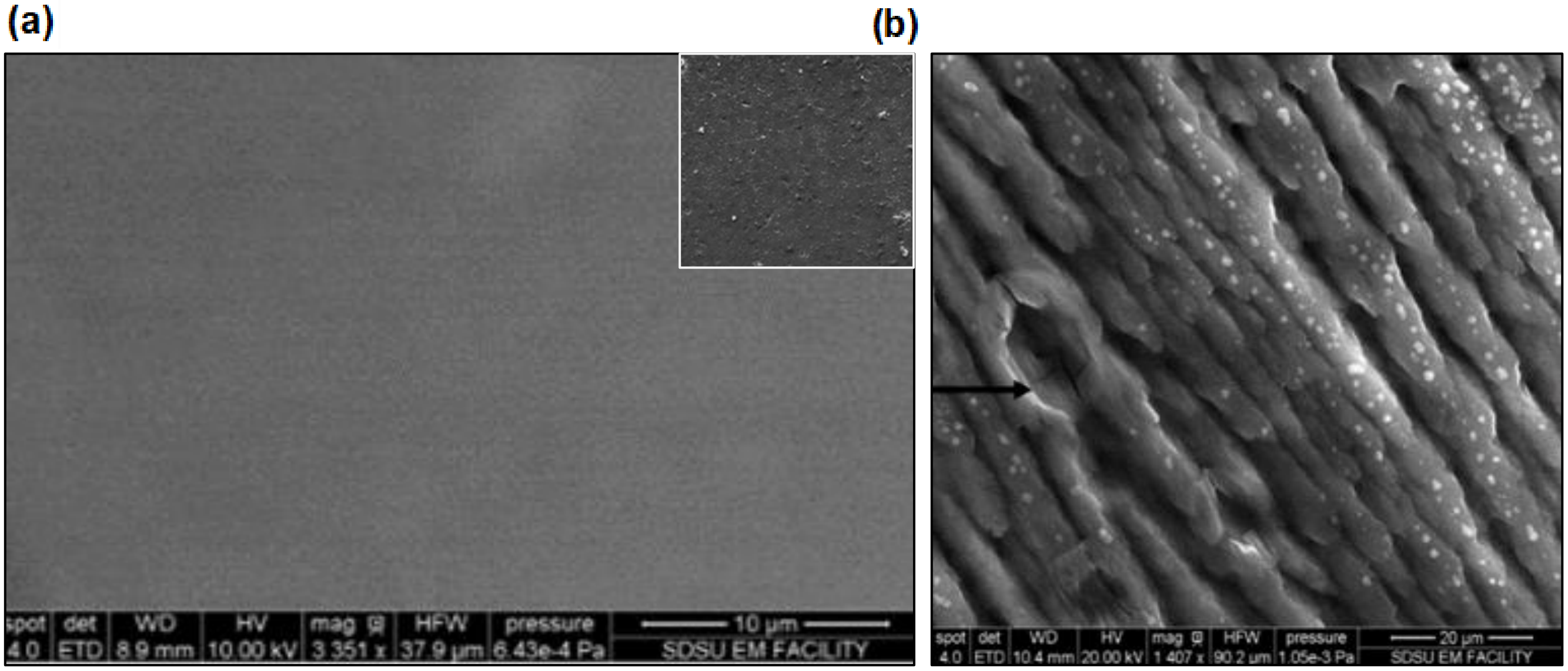

**Figure 3.** SEM image of (a) bare GC and (b) GC covered with chitosan matrix coating. The inset shows a high-resolution image of GC.

### 3.2. Electrical and Electrochemical Characterizations

Cyclic voltammetry (CV) results demonstrated stable and reproducible redox behavior across tested electrodes. The characteristic oxidation and reduction peaks observed in PBS confirm functional charge transfer at the GC interface. Minor pre-peaks detected near −0.2 V are likely associated with surface heterogeneity or fabrication-related micro-defects rather than intrinsic electrode material behavior (**Figure 4**). Electrochemical impedance spectroscopy (EIS) revealed relatively high charge-transfer resistance and non-ideal capacitive behavior across electrodes (**Figure 5**). Nyquist plots displayed semicircular features at low frequencies and diffusion-associated linear tails at higher frequencies. Bode plots indicated elevated impedance at low frequencies and phase shifts consistent with distributed capacitance.

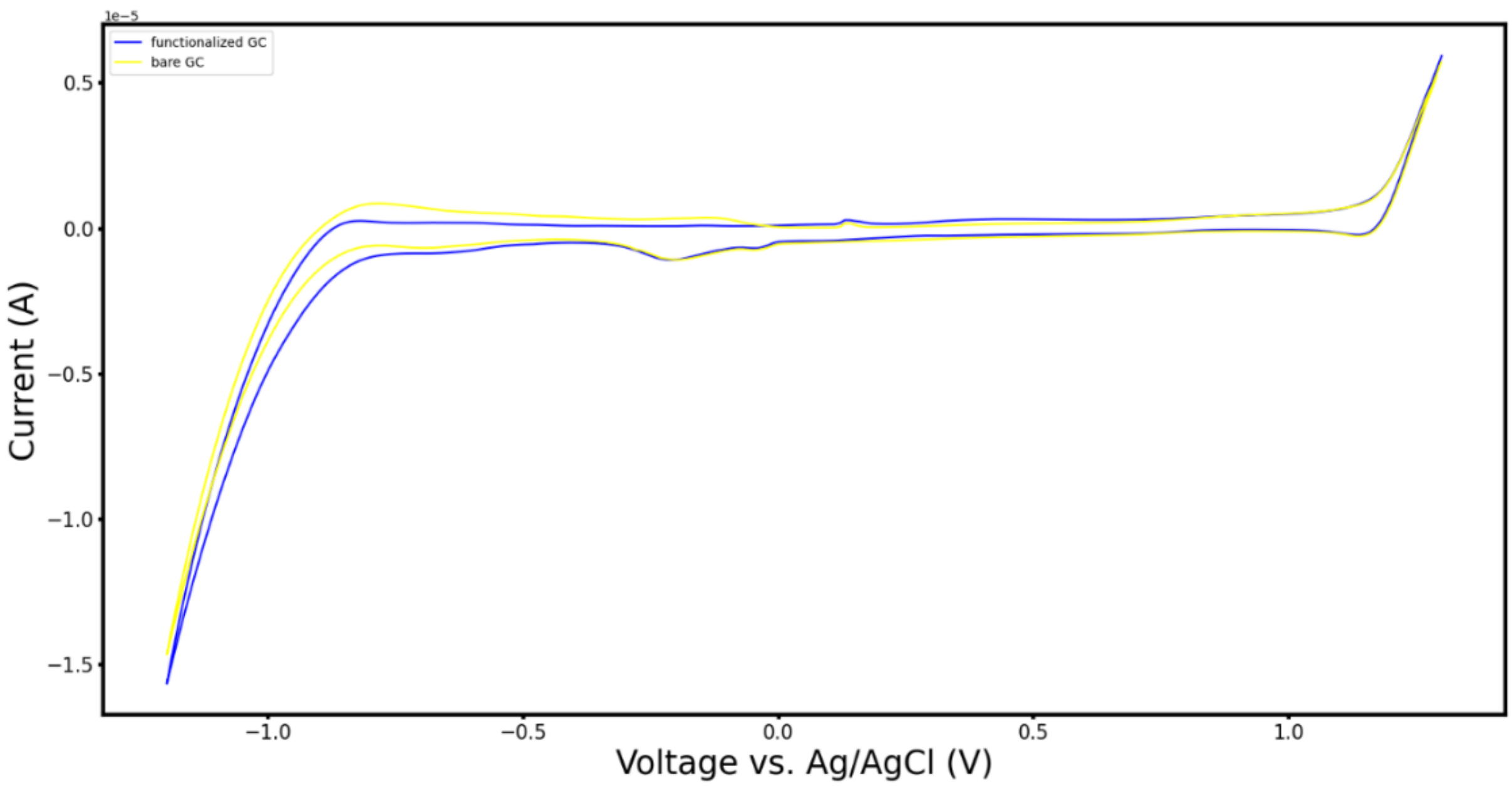

**Figure 4.** Cyclic voltammetry (CV) of the functionalized and bare GC microelectrodes of a 1.7 cm long probe with microelectrodes of 40 μm in diameter.

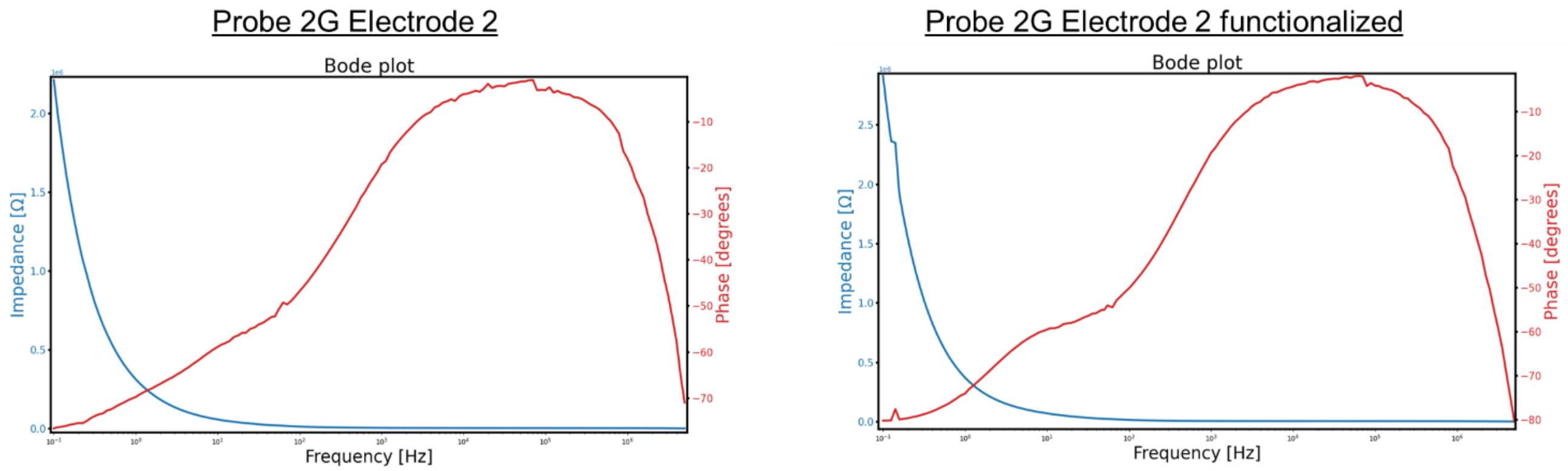

**Figure 5.** Bode plot for the functionalized and bare GC microelectrodes of the same neural probe.

### 3.3 FTIR Characterization

**Figure 6** summarizes the Fourier-transform infrared (FTIR) analysis of GC microelectrodes in their bare state, plasma-etched with a chitosan coating, with a chitosan coating alone, and a coated sample after mechanical washing. Initial analysis of the bare and plasma-etched GC microelectrodes primarily exhibited aromatic *C=C* stretching features, with plasma etching introducing no additional detectable functional groups beyond these baseline carbon features. Following the deposition of the chitosan matrix, significant spectral changes were observed. Both the etched and unetched coated samples (**Figure 6b, c**) displayed a prominent, broad absorption band between *3276* and *3414 $cm^{-1}$*. This breadth is attributed to the averaging of multiple *hydroxyl* (*O–H*) and *amine* (*N–H*) groups bonded to slightly different extents, creating a composite peak rather than isolated sharp signals [**33-34**]. Further characteristic chitosan markers were identified, including *Amide I* (*1635 $cm^{-1}$*), *Amide III* (1338 $cm^{-1}$), $CH_2$ bending (1406 $cm^{-1}$), and *C–O* stretching (1018 $cm^{-1}$). While $CH_3$ bending features were expected near 1375 $cm^{-1}$, they were likely obscured by the adjacent $CH_2$ peak. These results align closely with established literature values for chitosan, confirming the successful deposition of the enzymatic polymer matrix. In genral, the broad peak observed in the *3200–3500 $cm^{-1}$* region is a recognized feature of chitosan, representing the overlapping stretching vibrations of both hydroxyl (*O–H*) and amine (*N–H*) groups, which are further broadened by intramolecular hydrogen bonding [**35-36**]. Additionally, the specific *Amide I* and *Amide III* markers (typically appearing around *1635–1650 cm⁻¹* and *1313–1338 cm⁻¹,* respectively) are standard indicators used to confirm the presence of *N-acetyl* groups in the polymer matrix [**36-37**]. To evaluate the mechanical stability of the interface, the chitosan-coated samples were subjected to an agitated wash in deionized water for five minutes. As shown in **Figure 6d**, the FTIR spectra collected after washing and air-drying retained all characteristic chitosan peaks, including the signature *O–H* stretch and *amide* bands. The preservation of these spectral features indicates that the chitosan layer remains firmly adhered to the glassy carbon surface, demonstrating that the bond can withstand mechanical agitation under neutral pH conditions. **Figure 7** provides a side-by-side direct comparison of the FTIR results from all four samples.

### 3.3 Electrochemical Validation and Indirect Lactate Detection

#### 3.3.1 Hydrogen Peroxide Detection

Because lactate oxidase catalyzes the conversion of lactate to pyruvate and hydrogen peroxide (L-lactate + $O_2$ → pyruvate + $H_2O_2$), the ability of the GC neural probe to detect $H_2O_2$ was first validated. Then, using fast-scan cyclic voltammetry (FSCV), bare GC microelectrodes demonstrated reliable and concentration-dependent detection of hydrogen peroxide over a range of 10 nM to 2 μM. The oxidation peak occurred near +1.2 V vs Ag/AgCl and increased consistently with concentration, confirming effective electrochemical transduction. After chitosan coating, hydrogen peroxide detection was repeated to evaluate the effect of the polymer layer. Although a modest reduction in peak current amplitude was observed relative to bare GC, the coated electrodes retained clear and reproducible responses across the full concentration range. The reduced amplitude is attributed to diffusional resistance introduced by the polymer matrix rather than impaired electron transfer at the GC interface.

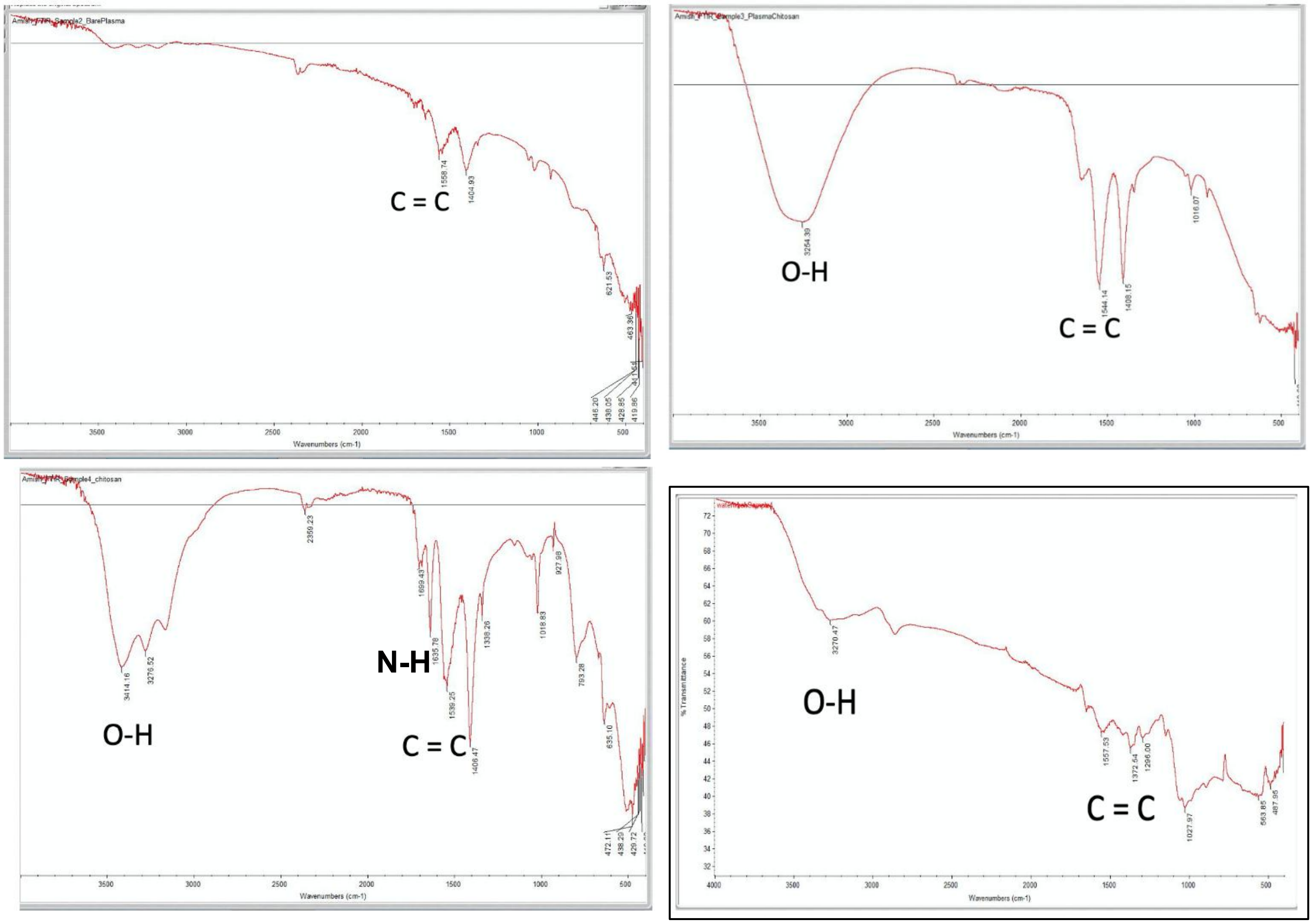

**Figure 6.** FTIR analysis of GC microelectrode (a) bare, (b) with plasma etching and chitosan coating and (c) with chitosan coating. (d) sample four FTIR analysis after a 5-minute agitated mechanical wash in DI water.

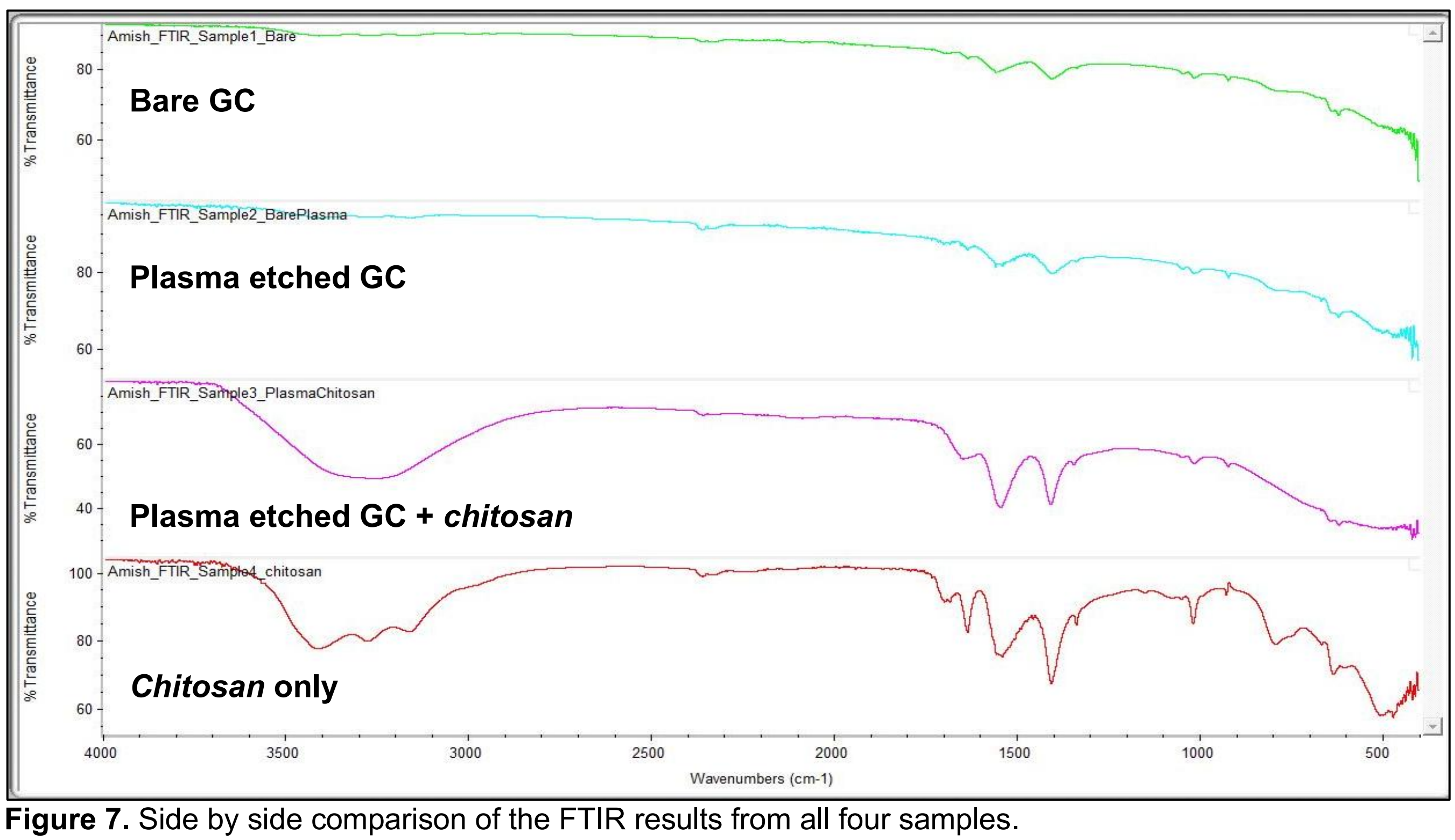

**Figure 7.** Side by side comparison of the FTIR results from all four samples.

### 3.3.2 Control Experiments: Absence of Direct Lactate Oxidation

**As shown in Figure 8a**, to confirm that lactate does not undergo direct electrooxidation under the applied FSCV waveform, unfunctionalized GC electrodes were exposed to lactate solutions under identical conditions. No measurable redox peaks were observed, verifying that lactate detection requires enzymatic conversion. Overlaying responses from functionalized and unfunctionalized electrodes following injection of 10 nM lactate demonstrated a clear increase in current only in the presence of immobilized lactate oxidase, confirming enzymatically mediated detection. As a control on chitosan coating, FSCV was carried out using CFM probe that was coated with chitosan. As shown in **Figure 8b**, lactate was successfully detected at concentrations from 10 nM – 2 μM.

### 3.3.3 *In Vitro* Lactate Detection Through FSCV

Fully functionalized GC microelectrodes were then evaluated for indirect lactate detection. Lactate concentrations ranging from 10 nM to 2 μM were introduced into phosphate-buffered saline, and the resulting current responses were recorded. Lactate was successfully detected at concentrations as low as 10 nM (**Figure 9**). However, the current response did not scale proportionally with concentration. For example, injection of 30 nM lactate generated a peak current of approximately 6.6 nA, whereas 1 μM and 2 μM produced peak currents of 6.9 nA and 8.0 nA, respectively. Despite more than an order-of-magnitude increase in concentration, the corresponding increase in current was modest. This nonlinear behavior suggests that signal generation is limited by enzymatic turnover rather than electrochemical transduction. At elevated substrate concentrations, lactate oxidase likely approaches catalytic saturation, resulting in a regime in which hydrogen peroxide production becomes constrained by the maximum reaction rate. Additional contributions from diffusion limitations within the chitosan matrix and oxygen availability may further restrict proportional scaling at higher concentrations.

### 3.3.4 Limit of Detection

The limit of detection (LOD) was calculated using the standard criterion, LOD = 3σ / slope, where σ represents the standard deviation of baseline noise and slope is derived from the linear portion of the calibration curve (10 nM – 2 μM, n = 7). As shown in **Figure 10**, the calculated theoretical LOD of 0.42 nM was within the low nanomolar range for both hydrogen peroxide and lactate detection under in vitro conditions. It should be noted that this LOD remains theoretical. In addition, while statistical detectability is in the nanomolar range, the practical quantitative dynamic range of lactate detection is constrained by enzyme-limited kinetics, as described above.

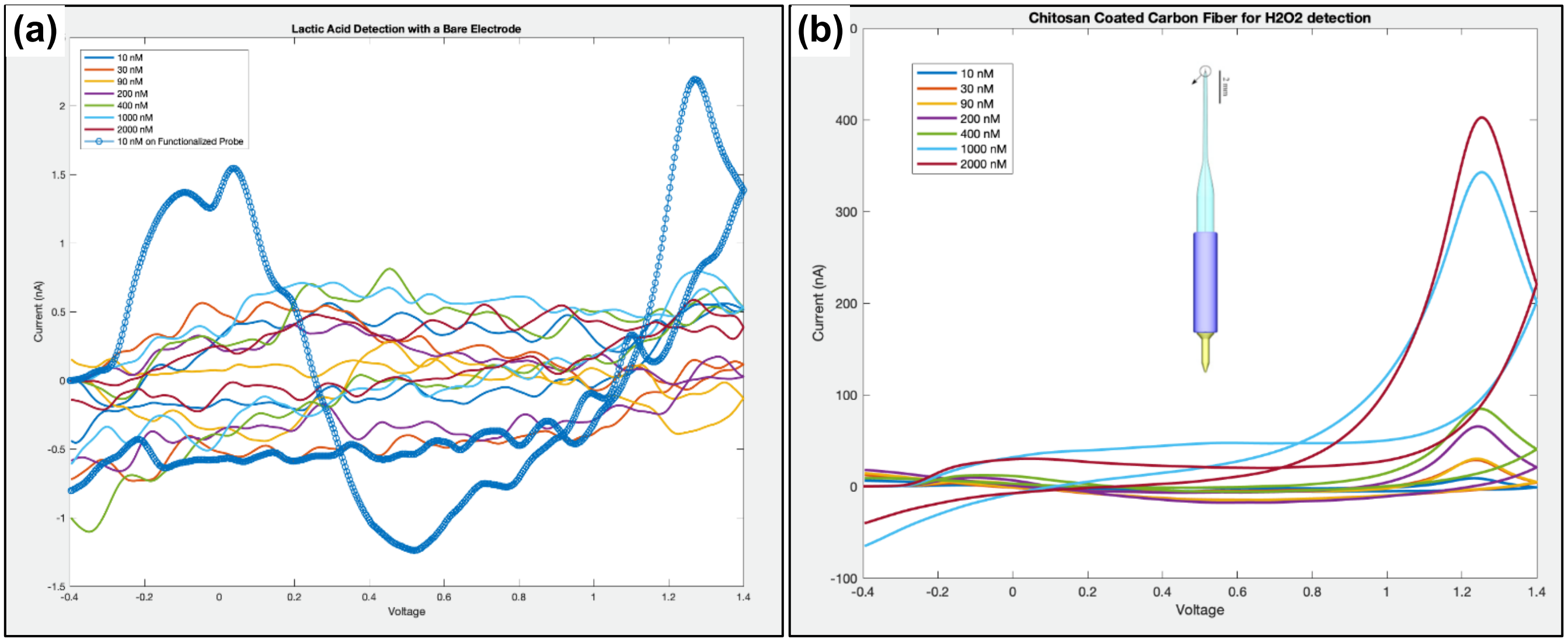

**Figure 8. Control experiments** (a) no redox reaction occurring on bare GC electrode; however, 10 nM of lactate in functionalized GC electrode shows a response, (b) carbon fiber probe coated with *chitosan* $H_2O_2$ detection.

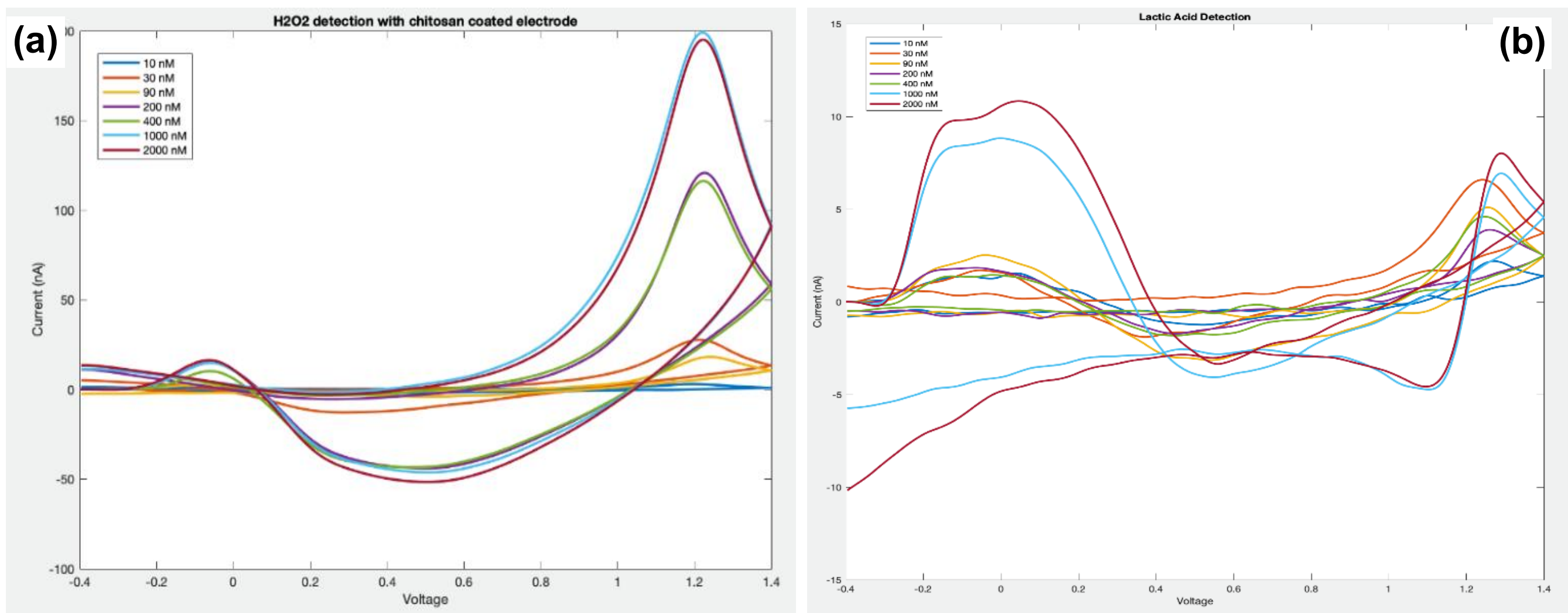

**Figure 9. (a)** Chitosan coated GC electrode was able to detect electrochemically active hydrogen peroxide (b) Lactate detection with fully functionalized GC microelectrodes.

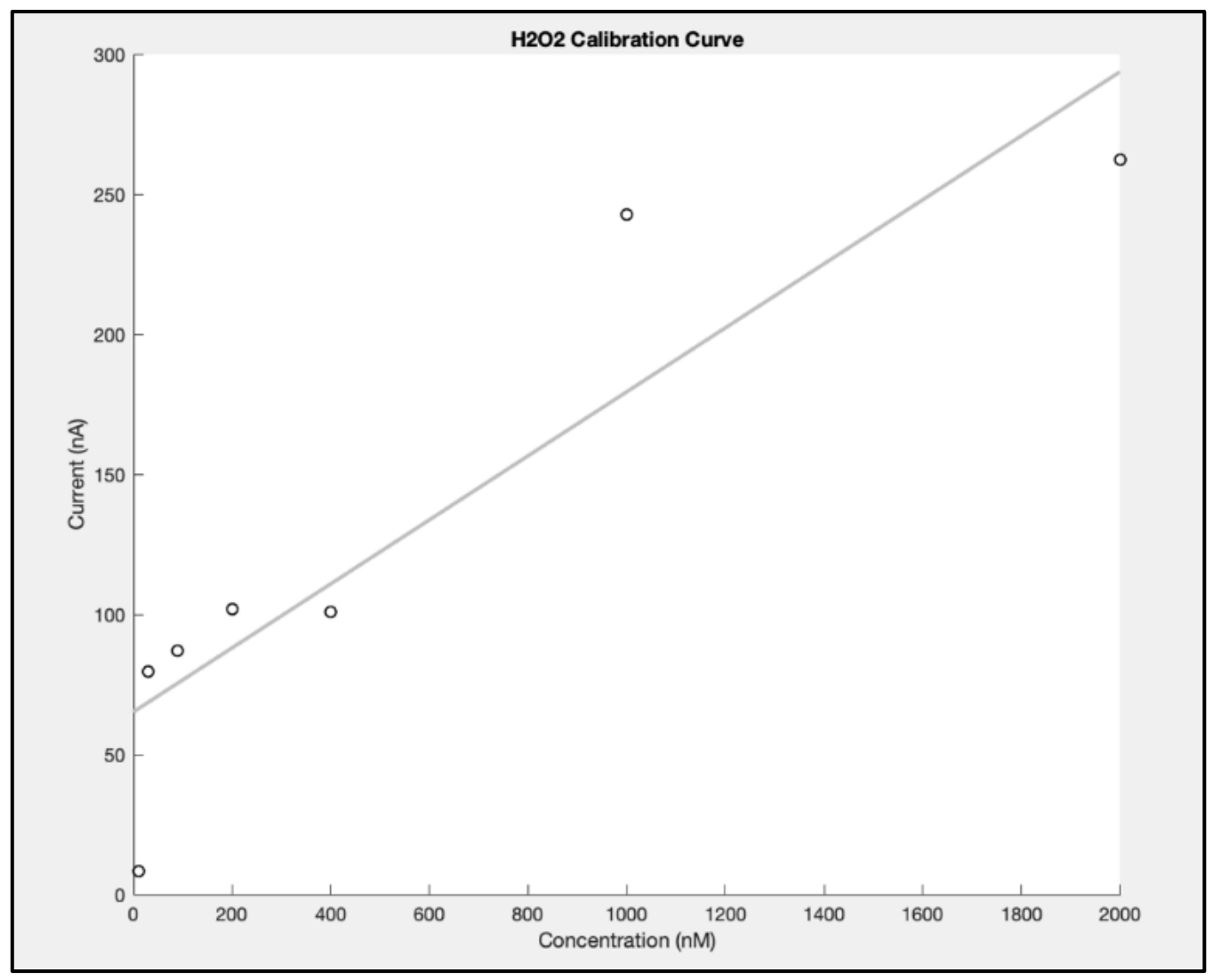

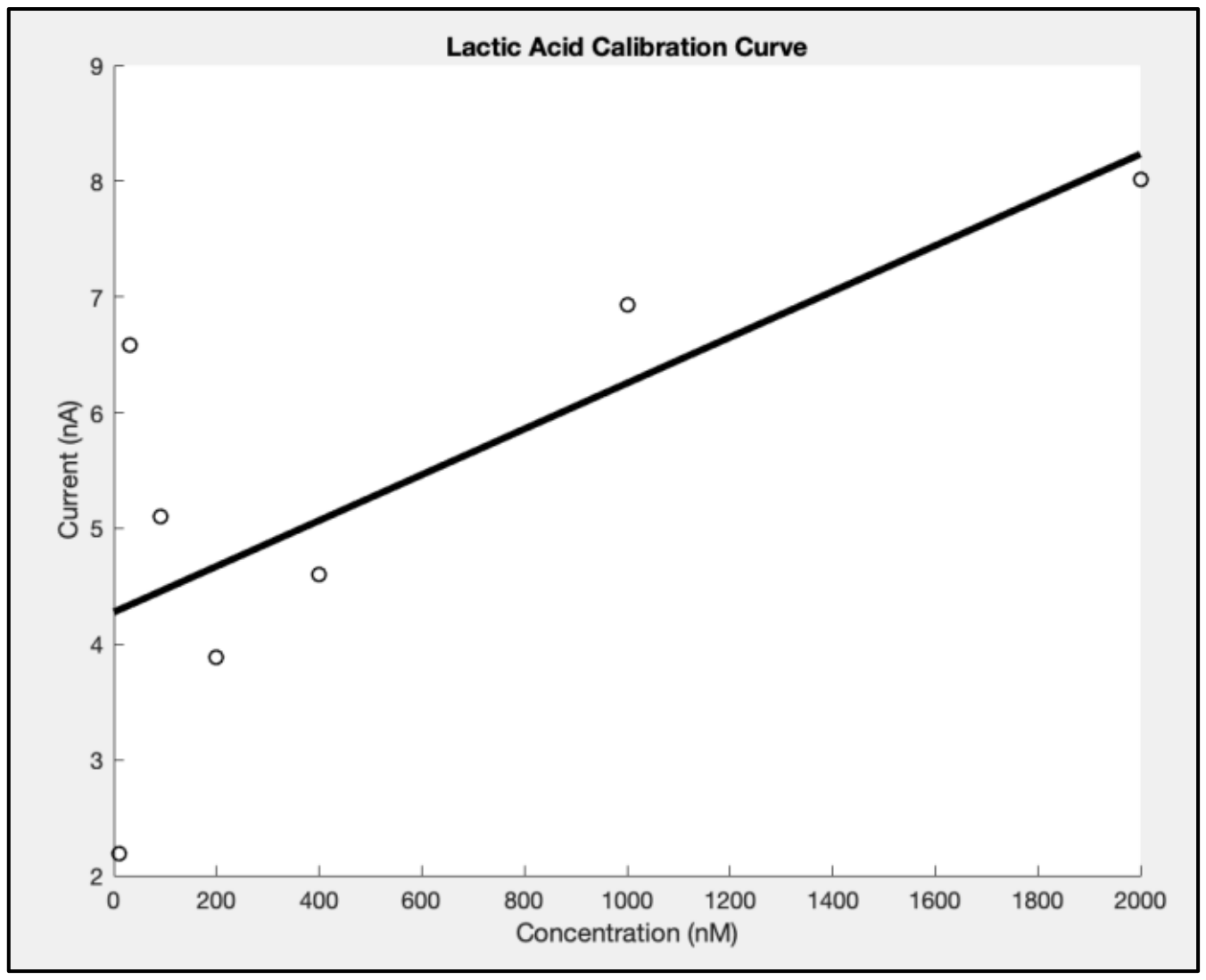

**Figure 10. LOD** (a) 0.42 nM for **c**arbon fiber probe coated with *chitosan* $H_2O_2$ detection (b) lactate calibration curve

## 4. Discussion

### 4.1 Electrochemical Transduction and Interfacial Charge Transfer

The glassy carbon microelectrode platform demonstrated robust hydrogen peroxide oxidation at ~+1.2 V vs Ag/AgCl under FSCV conditions. The oxidation response was concentration-dependent over 10 nM –2 μM, indicating that the applied waveform and GC surface provide sufficient overpotential and rapid heterogeneous electron transfer for efficient peroxide oxidation within this range.

Importantly, peroxide sensitivity was largely preserved after chitosan deposition. Although a modest reduction in peak current amplitude was observed, this attenuation is consistent with added diffusional resistance rather than impaired electron-transfer kinetics. Given the intrinsically favorable electron-transfer properties of GC toward peroxide oxidation, the dominant impedance introduced by the coating is mass-transport–controlled rather than charge-transfer–controlled. The absence of measurable response at bare electrodes upon lactate exposure confirms that detection is strictly enzyme-mediated. Thus, lactate sensing must be interpreted as a coupled enzymatic–electrochemical system in which electrochemical transduction is not the primary limiting factor.

As demonstrated in previous studies, GC facilitates high-resolution monitoring of bio-analytes and precise chemical signal detection. A critical driver of this high neurotransmitter sensitivity is the surface chemistry of the material; specifically, the intrinsic functional groups (like hydroxyl groups) that form on the GC surfaces and are ideal for detecting electrochemical activity. For example, these hydroxyl groups facilitate the adsorption of cationic species, such as dopamine, whose amine side chain becomes protonated at physiological pH [**38-39**].

### 4.2 Enzyme-Limited Lactate Detection and Reaction–Diffusion Coupling

Following immobilization of lactate oxidase within the chitosan matrix, indirect lactate detection was achieved at concentrations as low as 10 nM. However, the current response exhibited early saturation, with only modest increases in signals across orders-of-magnitude increases in substrate concentration. This compressed dynamic range indicates that signal generation is governed primarily by enzymatic turnover and mass transport within the immobilized layer rather than by electrochemical oxidation capacity. In enzyme-based electrochemical sensors, the measured current reflects a cascade of processes:

substrate diffusion into the polymer matrix, catalytic conversion at enzyme active sites, and diffusion of the reaction product to the electrode surface. When substrate supply exceeds catalytic throughput, additional increases in bulk concentration do not proportionally increase product flux. The early onset of apparent saturation suggests that reaction–diffusion coupling within the chitosan matrix constrains hydrogen peroxide generation. Restricted lactate and oxygen transport through the polymer networks are likely to reduce effective substrate availability at enzyme active sites. Under these conditions, the observed kinetic ceiling reflects transport limitations within the film rather than intrinsic enzyme affinity or electrode performance.

### 4.3 Influence of Polymer Microstructure

SEM imaging revealed a porous, interconnected chitosan morphology that provides mechanical stability and enzyme retention while introducing diffusion tortuosity. The modest attenuation in peroxide response following coating supports the presence of a diffusion barrier that becomes more influential in enzymatic sensing mode, where substrate transport precedes product formation.

Accordingly, polymer thickness, porosity, enzyme loading density, and oxygen permeability collectively define the quantitative dynamic range of lactate detection. Optimization strategies should focus on engineering the enzyme microenvironment—reducing film thickness, increasing porosity, or modulating enzyme density—to enhance catalytic throughput while preserving nanomolar sensitivity.

### 4.4 Analytical Performance and Implications

The calculated limits of detection fall within the low nanomolar range under in vitro conditions, comparable to FSCV detection of electroactive neurotransmitters using GC microelectrodes. Electrochemical sensitivity and temporal resolution remain intact after enzyme functionalization, demonstrating that the GC interface is not the performance bottleneck. Although lactate detection is constrained by enzyme-limited kinetics, the platform provides key attributes for neurochemical sensing: nanomolar detectability, rapid temporal response, stable polymer adhesion, and compatibility with lithographically patterned multi-site arrays. Future *in vivo* implementation will require evaluation of interferents at elevated potentials, enzyme stability, and oxygen availability under physiological conditions. Nonetheless, the present findings establish that catalytic throughput and substrate transport—not electrochemical transduction—govern sensor behavior in this architecture.

### 4.5 Mechanistic Summary

Collectively, the results demonstrate that glassy carbon enables efficient nanomolar peroxide oxidation and chitosan immobilization preserves interfacial charge transfer. The results also demonstrate that lactate detection is governed by enzyme kinetics and mass transport and dynamic range compression arises from reaction–diffusion coupling rather than electrode limitations. Accordingly, future performance improvements should prioritize engineering the enzyme microenvironment rather than modifying the electrochemical platform.

## 5. Conclusions

This work establishes a lithographically patterned glassy carbon (GC) microelectrode platform for indirect nanomolar detection of lactate via lactate oxidase–mediated hydrogen peroxide generation under fast-scan cyclic voltammetry. The results demonstrate that GC microelectrodes maintain robust electrochemical transduction following chitosan-based enzyme immobilization, preserving rapid temporal response and efficient peroxide oxidation. A key finding of this study is the decoupling of electrochemical transduction from enzymatic performance. While hydrogen peroxide detection remained concentration-dependent across the tested range, lactate sensing exhibited early saturation. This behavior indicates that analytical limitations arise not from the GC electrode interface, but from reaction–diffusion coupling and catalytic turnover within the immobilized enzyme matrix. Thus, dynamic range compression is governed by substrate transport and enzymatic throughput rather than heterogeneous electron-transfer kinetics.

These findings yield several important conclusions:

1. **Glassy carbon provides a mechanically robust and electrochemically stable platform suitable for enzyme-based neurochemical sensing.**
   Lithographic patterning enables scalable, multi-site integration while preserving nanomolar sensitivity.

2. **Enzyme immobilization within Chitosan preserves charge-transfer efficiency but introduces transport-mediated constraints.**
   The dominant impedance arises from diffusion within the polymer matrix rather than from electrochemical limitations at the electrode surface.

3. **Analytical performance in indirect lactate sensing is governed by enzyme kinetics and mass transport.**
   Consequently, optimization efforts should prioritize engineering the enzyme microenvironment—through control of film thickness, porosity, oxygen accessibility, and enzyme loading density.

Beyond demonstrating nanomolar *in vitro* detectability, this study clarifies the mechanistic origin of performance bottlenecks in enzyme-functionalized GC systems. By identifying catalytic and transport limitations as the primary governing factors, the work provides a framework for rational sensor optimization. Importantly, the integration of enzyme-functionalized GC microelectrodes into flexible, lithographically defined neural probes establishes a scalable foundation for multiplexed detection of electroactive neurotransmitters and metabolically relevant analytes within a unified device architecture [**40**]. Such capability is critical for advancing closed-loop neuromodulation, metabolic monitoring, and real-time neurochemical mapping in future in vivo applications.

**Author Contributions: AR** fabricated devices, implemented the microfabrication process, analyzed the results and wrote the materials and methods and results sections of the paper**; EB** fabricated devices, implemented the microfabrication process, did the functionalization, analyzed the results and wrote the materials and methods and results sections of the paper; **NR** helped in microfabrication; **SLG** helped in surface functionalization; **SN** helped in microfabrication and helped with SEM imaging; **M.M** helped with equipment and software set-up; **SO** helped with plotting results; and **SK** formulated the concept, supervised the project, structured the outline of the paper, edited the manuscript, and wrote the introduction, discussion and conclusion section of the paper.

**Funding:** This research is funded by the Center for Neurotechnology (CNT), a National Science Foundation Engineering Research Center (EEC-1028725) and NSF AccelNet: Broadening Carbon Ring program (Award Number: 2301898).

**Acknowledgment:** We would like to extend our gratitude to Dr. David Pullman for providing access to the FTIR spectrometer.

**Data Availability Statement:** Data is available upon request.

**Conflict of Interest:** The authors declare no competing financial interests.

**References**

1. Mittal, S., et al. (2023). The Impact of Neurotransmitters on the Neurobiology of Neurodegenerative Diseases. *International Journal of Molecular Sciences*.

2. Gründer, G., & Cumming, P. (2016). "*The Dopamine Hypothesis of Schizophrenia: Current Status*," In T. Abel & T. Nickl-Jockschat (Eds.), The Neurobiology of Schizophrenia (pp. 109–124). Elsevier Academic Press. https://doi.org/10.1016/B978-0-12-801829-3.00015-X
3. Ghanemi, A. (2013). Schizophrenia and Parkinson's disease: Selected therapeutic advances beyond the dopaminergic etiologies. *Alexandria Journal of Medicine*. J. Lotharius and P. Brundin, Nature Rev Neurosci, 3(12), 932-942 (2002).
4. Koob, G. F., & Volkow, N. D. (2016). Neurobiology of addiction: a neurocircuitry analysis. *The Lancet Psychiatry*.
5. Panigrahi B, Martin KA, Li Y, Graves AR, Vollmer A, et al. 2015. "*Dopamine Is Required for the Neural Representation and Control of Movement Vigor*," *Cell* 162: 1418-30
6. Dunigan AI, Roseberry AG., "*Actions of Feeding-Related Peptides On the Mesolimbic Dopamine System in Regulation of Natural and Drug Rewards," Addict Neurosci* 2, 2022.
7. Wise, R. A., & Robble, M. A. (2020). *Dopamine and Addiction*. *Annual Review of Psychology*.
8. Luscher, C. (2016). *The Emergence of a Circuit Model for Addiction*. *Annual Review of Neuroscience*.
9. Barros, L. F. (2013). *Metabolic signaling by lactate in the brain*. *Trends in Neurosciences*.
10. O. Matz, C. Zdebik, S. Zechbauer, L. Bündgens, J. Litmathe, K. Willmes, J.B. Schulz, M. Dafotakis,"*Lactate as a diagnostic marker in transient loss of consciousness*," Seizure, Volume 40, 2016, Pages 71-75, ISSN 1059-1311, https://doi.org/10.1016/j.seizure.2016.06.014.
11. Chen X, Zhang Y, Wang H, Liu L, Li W, Xie P. *The regulatory effects of lactic acid on neuropsychiatric disorders*. Discov Ment Health. 2022 Mar 30;2(1):8. doi: 10.1007/s44192-022-00011-4. PMID: 37861858; PMCID: PMC10501010.
12. Wang, J., et al. (2023). From metabolic substrate to epigenetic regulation: roles and mechanisms of lactylation in brain health and disease. *Frontiers in Molecular Neuroscience*.
13. Chen, P., et al. (2024). Dual roles of lactate and lactylation modification in the nervous system: neuroprotection and neuroinjury. *Frontiers in Aging Neuroscience*.
14. Zhou, X., et al. (2024). Lactate and cognition: a dual modulator. *Frontiers in Physiology*.
15. Dienel, G. A. (2023). Brain Lactate Dynamics: The Polyfunctional Role of Lactate in Brain Metabolic Strategy and Signaling. *Journal of Neuroscience Research*.
16. Ou, Y., et al. (2019). *Frontiers in electrochemical sensors for neurotransmitter detection: Towards measuring neurotransmitters as chemical diagnostics for brain disorders*. *Analytical Methods*.
17. Bucher, E. S., & Wightman, R. M. (2015). High-Resolution Spatiotemporal Measurements of Neurochemistry. *Annual Review of Analytical Chemistry*.
18. Robinson, D. L., et al. (2003). Detecting Subsecond Dopamine Release with Fast-Scan Cyclic Voltammetry in Free-Moving Rats. *Clinical Chemistry*.
19. Ventura, R. D., et al. (2023). Electrochemical Sensing of Neurotransmitters: A Comprehensive Review of Materials and Methods. *Chemical Reviews*.
20. Lakard, B. (2020). Electrochemical Detection of Neurotransmitters. *Sensors*.
21. Galindo, Sandra Lara, Surabhi Nimbalkar, Alexis Oyawale, James Bunnell, Omar Nunez Cuacuas, Rhea Montgomery-Walsh, Amish Rohatgi, Brinda Kodira Cariappa, Abhivyakti Gautam, Kevin Peguero-Garcia, and et al. 2024. "*Indirect Voltammetry Detection of Non-Electroactive Neurotransmitters Using Glassy Carbon Microelectrodes: The Case of Glutamate*" *C* 10, no. 3: 68. https://doi.org/10.3390/c10030068
22. Whulanza Y, Arafat Y, Rahman S, et al. "*On-chip testing of a carbon-based platform for electro-adsorption of glutamate*", Heliyon, 2022; 8
23. Dale, N., Hatz, S., Tian, F., & Llaudet, E. (2005). "*Real-time monitoring of neurochemical systems in vivo: The role of enzyme-based biosensors," Biological Psychiatry*.
24. Wilson, G. S., & Gifford, R. (2005). Biosensors for real-time in vivo measurements. *Biosensors and Bioelectronics*.
25. Ross, A. E., & Venton, B. J. (2012*). Sawhorse Waveform Can Be Used To Measure Adenosine and Dopamine Simultaneously with Fast-Scan Cyclic Voltammetry*. *Analytical Chemistry*.
26. Skoog, G. L., & Wightman, R. M. (2013). Enzyme-Modified Carbon-Fiber Microelectrode for the Quantification of Dynamic Fluctuations of Non-Electroactive Analytes Using Fast-Scan Cyclic Voltammetry. *Analytical Chemistry*.

27. Zestos, A. G., et al. (2018). Carbon Nanomaterials for Electrochemical Sensing. *ChemElectroChem*.
28. Castagnola, E., Vahidi, N. W., Nimbalkar, S., & Rudraraju, S. (2018). "*In Vivo Dopamine Detection and Single Unit Recordings Using Intracortical Glassy Carbon Microelectrode Arrays*." MRS Advances, 3(29), 1629–1634. https://doi.org/10.1557/adv.2018.98.
29. Castagnola, E, S Thongpang, M Hirabayashi, G Nava, S Nimbalkar, T Nguyen, S Lara, et al. 2021. "*Glassy Carbon Microelectrode Arrays Enable Voltage-Peak Separated Simultaneous Detection of Dopamine and Serotonin Using Fast Scan Cyclic Voltammetry*." Analyst 146 (12): 3955–70. https://doi.org/10.1039/d1an00425e.
30. Vomero, M., Van Niekerk, P., Nguyen, V., Gong, N., Hirabayashi, M., Cinopri, A., Logan, K., Moghadasi, A., Varma, P., and Kassegne. S., 2016. "*A Novel Pattern Transfer Technique for Mounting Glassy Carbon Microelectrodes on Polymeric Flexible Substrates*." Journal of Micromechanics and Microengineering 26 (2). doi.org/10.1088/0960-1317/26/2/025018.
31. Nimbalkar, S., Castagnola, E., Balasubramani, A., Scarpellini, A., Samejima, S., Khorasani, A., Boissenin, A., Thongpang, S., Moritz, C., and Kassegne, S., "*Ultra-Capacitive Carbon Neural Probe Allows Simultaneous Long-Term Electrical Stimulations and High-Resolution Neurotransmitter Detection"*, Nature Scientific Reports 8 (1), 6958, 2018**.**
32. Vomero, Maria, Elisa Castagnola, Francesca Ciarpella, Emma Maggiolini, Noah Goshi, Elena Zucchini, Stefano Carli, Luciano Fadiga, Sam Kassegne, and Davide Ricci. 2017. "*Highly Stable Glassy Carbon Interfaces for Long-Term Neural Stimulation and Low-Noise Recording of Brain Activity*." *Scientific Reports* 7 (December 2016): 1–14. https://doi.org/10.1038/srep40332.
33. Nandiyanto, A, R Oktiani, and R Ragadhita. 2019. "*How to Read and Interpret FTIR Spectroscope of Organic Material*." Indonesian Journal of Science and Technology 4 (1): 97–118. https://doi.org/10.17509/ijost.v4i1.15806.
34. Mureşan-Pop, M, I Kacsó, X Filip, E Vanea, G. Borodi, N. Leopold, I. Bratu, and S. Simon. 2011. "*Spectroscopic and Physical-Chemical Characterization of Ambazone-Glutamate Salt*." Spectroscopy 26 (2): 115–28. https://doi.org/10.3233/SPE-2011-0519.
35. Gieroba, B., Sroka-Bartnicka, A., Kazimierczak, P., Kalisz, G., Lewalska-Graczyk, A., Vivcharenko, V., Nowakowski, R., Pieta, I. S., & Przekora, A. (2022). "*Surface chemical and morphological analysis of chitosan/1,3-β-d-glucan polysaccharide films cross-linked at 90* °C," International Journal of Molecular Sciences, 23(11), 5953. https://doi.org/10.3390/ijms23115953
36. Kumirska, J., Czerwicka, M., Kaczyński, Z., Bychowska, A., Brzozowski, K., Thöming, J., & Stepnowski, P. (2010). Application of spectroscopic methods for structural analysis of chitin and chitosan. Marine Drugs, 8(5), 1567–1636. https://doi.org/10.3390/md8051567
37. Islam, S., Arnold, L., & Padhye, R. (2015). "*Comparison and Characterisation of Regenerated Chitosan from 1-Butyl-3-methylimidazolium Chloride and Chitosan from Crab Shells*," *BioMed Research International*, *2015*, 1–6. https://doi.org/10.1155/2015/874316
38. Montgomery-Walsh, R., Nimbalkar, S., Bunnell, J., Galindo, S.L., **Kassegne**, S., "*Molecular Dynamics Simulation of Evolution of Nanostructures and Functional Groups in Glassy Carbon under Pyrolysis,*" Carbon, Volume 184, **2021.**
39. Nimbalkar, S., Montgomery-Walsh, R., Bunnell, J., Galindo, S.L., Cariappa, B.K., Gautam, A., Arvizu, R., Yang, S., and **Kassegne**, S., "*Carbon Allotropes form a Hybrid Material: Synthesis, Characterization, and Molecular Dynamics Simulation of Novel Graphene-Glassy Carbon Hybrid Material"*, Carbon, **2022**.
40. Siwakoti, U., Jones, S. A., Kumbhare, D., Cui, X. T., & Castagnola, E. (2025). "*Recent progress in flexible microelectrode arrays for combined electrophysiological and electrochemical sensing*," Biosensors, 15(2), 100. https://doi.org/10.3390/bios15020100